\documentclass[runningheads]{svproc}

\usepackage{graphicx}
\usepackage{url}
\usepackage{booktabs}
\usepackage{multirow}
\usepackage{amsmath}
\usepackage{enumitem}
\usepackage{hyperref}
\usepackage{tikz}
\usetikzlibrary{arrows.meta,positioning,shapes.multipart,shadows} 
\usepackage{placeins}
\mainmatter             

\title{Improving Hospital Process Management through Process Mining:\\
A Case Study on COVID-19 Clinical Pathways}

\titlerunning{Process Mining for Hospital Management}

\author{Pasquale Ardimento\inst{1} \and Mario Luca Bernardi\inst{2} \and Marta Cimitile\inst{3} \and Samuele Latorre\inst{1}}
\authorrunning{Ardimento et al.} 
\institute{University of Bari Aldo Moro, Bari, Italy,\\
\email{pasquale.ardimento@uniba.it, s.latorre11@studenti.uniba.it}
\and
Unisannio University of Benevento, Italy \email{bernardi@unisannio.it}
\and
UnitelmaSapienza Rome, Italy \email{marta.cimitile@unitelmasapienza.it}}

\begin{document}
\maketitle

\begin{center}
\small
This is a preprint (submitted version) of a paper accepted for publication and presentation at WorldCIST 2026. 
This version has not undergone post-acceptance improvements, copy-editing, or typesetting. 
The Version of Record will be published in \textit{Lecture Notes in Networks and Systems} (Springer Nature). 
The final authenticated version will be available via its DOI.
\end{center}

\begin{abstract}
This study analyzes COVID-19 care pathways using the COVID Data for Shared Learning dataset. We build a transparent, reproducible pipeline that transforms heterogeneous clinical tables into a process-mining-ready event log and applies discovery, declarative conformance checking, and outcome analysis. The reconstructed pathways highlight the monitoring backbone of inpatient care, variability at the Emergency department-admission interface, and outcome differences driven by age and exposure to intensive care units.
These insights support triage standardization, capacity planning, and step-down coordination from intensive care units to lower-acuity wards, showing how process mining can inform evidence-based hospital governance.
\end{abstract}

\section{Introduction}
\label{sec:intro}
Hospitals must balance quality, safety, and efficiency of care. During the COVID-19 pandemic, Emergency Departments (ER) and Intensive Care Units (ICU) faced extreme variability in demand and severity, revealing the limits of traditional reporting for real-time process visibility. Understanding patient flows, delays, and pathway variability became critical for clinical and operational decision-making. \emph{Process mining} offers a data-driven means to reconstruct and analyze these flows from routinely collected data, revealing actual care trajectories, deviations from expected behavior, and their relationship with outcomes. However, many studies emphasize algorithms over the \emph{translational} value of process mining for hospital governance. This paper presents a case study on COVID-19 patient pathways using the \emph{COVID Data for Shared Learning (CDSL)} dataset \cite{Ritore2024CDSL} (PhysioNet v1.0.0) \cite{physioNet}, covering 4,479 patients from Dec~2019 to Feb~2021. We develop a reproducible pipeline that transforms clinical data into an event log and produces interpretable indicators, dominant routes, throughput times, and outcome gradients, relevant for triage and ICU coordination. We address three research questions:
\begin{enumerate}[label=RQ\arabic*,leftmargin=*,nosep]
  \item What are the dominant ER$\rightarrow$hospital$\rightarrow$discharge trajectories and where do bottlenecks emerge?
  \item How do real traces conform to a reference pathway, and which deviations are operationally meaningful?
  \item How do patient and process factors (age, ICU use, monitoring intensity) relate to discharge outcomes?
\end{enumerate}

Our contributions are:
\begin{enumerate}[label=C\arabic*,leftmargin=*,nosep]
  \item A documented pipeline for deriving hospital-grade event logs from restricted clinical data.
  \item An end-to-end analysis combining discovery, declarative conformance, and enhancement to support managerial interpretation.
  \item Empirical insights linking process evidence to governance levers for triage, capacity, and discharge.
\end{enumerate}

The paper is structured as follows. Section~\ref{sec:related} reviews related work; Section~\ref{sec:dataset} presents the dataset; Section~\ref{sec:methodology} outlines the methodology; Section~\ref{sec:results} reports findings; Section~\ref{sec:discussion} discusses implications; and Section~\ref{sec:conclusion} concludes.

\section{Related Work}
\label{sec:related}
Process mining has been widely used in healthcare to reconstruct care pathways, assess adherence to expected behavior, and measure performance variability across services and populations.  
Surveys and monographs highlight three core techniques, that are discovery, conformance checking, and enhancement, and emphasize that reliable results depend on sound \emph{data engineering}: defining activity semantics, reconciling timestamps, and identifying cases correctly \cite{Aalst2016,Rojas2016}. Despite these advances, most studies still emphasize algorithmic performance over transparent and reproducible data pipelines or managerial interpretation \cite{Rojas2016,Partington2015,MunozGama2022}. This work builds on prior efforts but places specific emphasis on transparent data transformation from raw clinical tables to an auditable event log suitable for both clinical and operational use.

Applications to emergency and acute-care settings reveal high pathway variability, critical decision points (triage, admission), and bottlenecks that cascade toward wards and ICUs.  
During the COVID-19 pandemic, process mining supported rapid understanding of patient trajectories, demand surges, and capacity planning \cite{Reinkemeyer2021,Augusto2022}.  
Both procedural and declarative representations have been adopted: directly-follows graphs for communicability, and Declare constraints for flexible yet analyzable behavior.  
Our study adopts this dual view, discovery for summarizing hospital flow and declarative conformance to highlight meaningful deviations, such as incomplete triage or atypical ER–admission branching.

In addition to clinical guideline compliance, process mining can inform operational decisions such as bed planning, ICU coordination, and resource allocation. Evidence-based indicators, rework, waiting time, or length of stay, become actionable when presented transparently to decision makers \cite{Mans2015,Giani2020}. 

Compared to the state of the art, our study contributes to: 1) document a reproducible pipeline from multimodal, restricted data to a hospital-grade event log; 2) integrate discovery and declarative conformance to produce interpretable, management-oriented insights; 3) link pathway evidence to outcome-based \emph{Key Performance Indicators} (KPIs) for triage, monitoring, and capacity governance.
\section{Dataset and Preparation}
\label{sec:dataset}

This study uses the \emph{CDSL} dataset, provided by HM Hospitales and distributed via PhysioNet (v1.0.0). The dataset includes 4,479 COVID-19 patients hospitalized between December 2019 and February 2021, with multimodal data covering demographics, ER and inpatient episodes, ICU stays, vital signs, laboratory tests, medications, diagnoses, and procedures.  
Access to CDSL requires credentialed authorization under the \emph{PhysioNet Contributor Review Health Data License~1.5.0} and a signed Data Use Agreement (DUA). All data were handled in full compliance with these conditions, and no re-identification or linkage beyond the authorized scope was attempted. The analysis included all patients present in the CDSL dataset who experienced at least one emergency or inpatient encounter with valid timestamps.  
In this dataset, each \texttt{patient\_id} corresponds to a single clinical trajectory, so no further segmentation of episodes was required. Therefore, all  records were treated as complete episodes of care, directly assigned to a unique \texttt{case\_id} in the event log.

\subsection{Integration and Harmonization Pipeline}
\label{subsec:integration}
Having defined the structure of the case, the next step was to merge the six source tables into a unified sequence of events, ordered in time.  
The raw data consist of six normalized CSV files provided by CDSL:

\begin{itemize}[leftmargin=*,nosep]
  \item \texttt{patient\_01.csv} — demographics and ER/inpatient/ICU timestamps,
  \item \texttt{diagnosis\_er\_02.csv} — ER diagnoses,
  \item \texttt{diagnosis\_hosp\_03.csv} — inpatient diagnoses,
  \item \texttt{vital\_signs\_04.csv} — measurements of vital parameters,
  \item \texttt{medication\_05.csv} — administered drugs,
  \item \texttt{lab\_06.csv} — laboratory test results.
\end{itemize}

These tables are linked through patient identifiers and timestamps to construct a chronologically ordered sequence of clinically meaningful events.  
A consolidated \textbf{activity label set} maps raw records to process activities covering ER admission, triage and discharge, ward and ICU transitions, and routine monitoring activities such as vitals, laboratory tests, and medication administrations.

Timestamp validation follows deterministic rules.  
When missing or incomplete timestamps resulted in an implausible or temporally inconsistent sequence (e.g., discharge preceding admission), hour values were imputed using the nearest clinically reliable timestamp for that patient.  
This correction was applied only in cases where it restored plausibility; otherwise, the affected records were kept unchanged. The resulting harmonized log preserves the temporal and clinical coherence required for discovery, conformance checking, and outcome analysis.

\subsection{Quality Control and Export}
Data quality checks addressed duplicates, negative durations, and temporally inconsistent sequences (e.g., discharge recorded before admission).  
Events violating temporal or clinical plausibility were not removed by default: when inconsistencies could be corrected through deterministic timestamp reconciliation (Section~\ref{subsec:integration}), the corrected version was retained.
The resulting event log includes three core columns, \texttt{case\_id}, \texttt{activity}, and \texttt{timestamp}, along with attributes related to the patient (age, sex, and previous admission details), and fields characterizing each activity.
The final log was exported in \textbf{XES format} in accordance with standard process mining practice. All transformations follow deterministic and fully documented rules to ensure methodological reproducibility, even though access to raw data remains restricted. 
Table~\ref{tab:logschema} lists all attributes extracted from the six CDSL source tables and made available during event-log construction. Only the attributes in \textbf{bold} were used in this study; the remaining attributes were retained for completeness but were not included in the analysis.

\subsection{Reproducibility}
The entire data transformation pipeline, from raw data ingestion to event log generation, is implemented in Python using Jupyter notebooks, and relies on the \texttt{pandas} and \texttt{PM4Py} libraries. All steps are fully scripted and reproducible, with no manual intervention required during execution. Once parameters such as mapping rules and timestamp logic are defined, the pipeline executes end-to-end deterministically across runs. Export to XES format is handled programmatically to ensure compatibility with standard process mining tools. All transformations are logged and governed by documented rules, supporting transparency, repeatability, and auditability.
In addition, the full pipeline implementation, along with documentation of the source tables, extracted fields, and activity mapping rules, is publicly available at https://github.com/HimThisGuy/CDSLtoEventLog

\section{Methodology}
\label{sec:methodology}
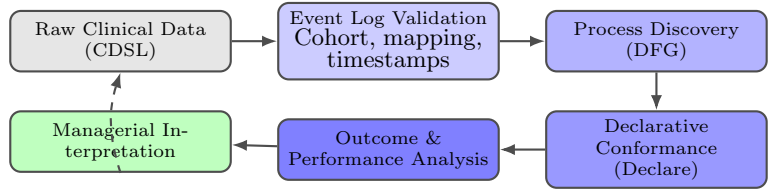
\begin{figure}[t]
\centering
\begin{tikzpicture}[
  every node/.style={font=\scriptsize, align=center},
  process/.style={rectangle, rounded corners=4pt, text width=2.7cm, minimum height=8mm,
                  draw=black!70, thick, fill=#1},
  arrow/.style={-{Latex[length=2mm]}, thick, black!70},
  node distance=5mm and 6mm
]
\node[process=gray!20]                  (data) {Raw Clinical Data\\(CDSL)};
\node[process=blue!20, right=of data]   (prep) {Event Log Validation\\\footnotesize Cohort, mapping, timestamps};
\node[process=blue!30, right=of prep]   (disc) {Process Discovery\\(DFG)};

\node[process=green!25,  below=of data] (mgmt) {Managerial Interpretation};
\node[process=blue!50,   below=of prep] (enh)  {Outcome \&\\Performance Analysis};
\node[process=blue!40,   below=of disc] (conf) {Declarative Conformance\\(Declare)};

\draw[arrow] (data) -- (prep);
\draw[arrow] (prep) -- (disc);
\draw[arrow] (disc) -- (conf);   
\draw[arrow] (conf) -- (enh);    
\draw[arrow] (enh)  -- (mgmt);   

\draw[arrow, dashed, bend left=18] (mgmt.south) to (data.south);
\end{tikzpicture}
\caption{Conceptual pipeline of the proposed methodology.}
\label{fig:workflow}
\end{figure}

This study adopts a structured process mining approach to extract clinically relevant insights from hospital data. To reflect the flexibility of real-world care pathways, especially in emergency and critical care, a declarative modeling perspective was chosen. The defined pipeline, shown in Figure \ref{fig:workflow}, is detailed below.

\begin{table}[ht]
    \centering
    \begin{tabular}{@{}l@{\hspace{2em}}l@{\hspace{2em}}l@{}}
    \toprule
    \textbf{Column} & \textbf{Type}  & \textbf{Source} \\
    \midrule
     \textbf{case\_id}                   & integer                   & any table              \\
     \textbf{activity}                   & categorical               & mapped                 \\
     \textbf{timestamp}                  & datetime                  & mixed                  \\
    \midrule
     \textbf{age}                        & integer                   & patient\_01            \\
     \textbf{sex}                        & categorical               & patient\_01            \\
     \textbf{previous\_admission\_date}  & datetime                  & patient\_01            \\
     \textbf{previous\_diagnosis}        & string                    & patient\_01            \\
    \midrule
     \textbf{er\_admission\_diagnosis}   & string                    & patient\_01            \\
     er\_department                      & string                    & patient\_01            \\
    \midrule
     blood\_press\_max                   & integer                   & patient\_01 \& vital\_signs\_04 \\
     blood\_press\_min                   & integer                   & patient\_01 \& vital\_signs\_04 \\
     temp                                & float                     & patient\_01 \& vital\_signs\_04 \\
     heart\_rate                         & integer                   & patient\_01 \& vital\_signs\_04 \\
     O2\_saturation                      & integer                   & patient\_01 \& vital\_signs\_04 \\
     glucose                             & integer                   & patient\_01 \& vital\_signs\_04 \\
    \midrule
     \textbf{er\_destination}            & string                    & patient\_01           \\
     er\_principal\_diagnosis            & string                    & diagnosis\_er\_02     \\
    \midrule
     \textbf{hosp\_admission\_diagnosis} & string                    & patient\_01           \\
    \midrule
     lab\_number                         & string                    & lab\_06               \\
    \midrule
     \textbf{hosp\_destination}          & categorical               & patient\_01           \\
     principal\_diagnosis                & string                    & diagnosis\_hosp\_03   \\
     principal\_diagnosis\_poa           & categorical               & diagnosis\_hosp\_03   \\
    \bottomrule
    \end{tabular}
    \caption{Attributes extracted from the six CDSL source tables.}
\label{tab:logschema}
\end{table}

\textbf{Step 1. Event Log Validation.}
The CDSL-derived event log (Section~\ref{sec:dataset}) was validated for completeness, timestamp coherence, and consistency across ER, ward, and ICU events.  
Each \texttt{case\_id} corresponds to a single patient, as the dataset provides at most one care pathway per patient.  
When implausible temporal sequences were detected (e.g., events occurring after discharge or timestamps lacking an hour specification), corrections were applied using the closest clinically valid timestamp rather than discarding events.  
Activity labels were checked against a predefined label set to ensure a consistent and non-redundant mapping from raw clinical fields to process activities.

\textbf{Step 2. Process Discovery.}
Process discovery was conducted using the Disco tool, which constructs a directly-follows model from the validated event log.  
The resulting model shows the dominant ER$\rightarrow$Hospital Stay$\rightarrow$Discharge trajectory and optional transitions to ICU.  
Since patients have at most one ER encounter in the dataset, no ER revisit loops appear.  
Frequency-based filtering was used to highlight representative transitions while preserving clinical interpretability.

\textbf{Step 3. Declarative Conformance Checking.}
A declarative reference model was defined in the \emph{Declare} language~\cite{Pesic2007}, encoding clinically meaningful constraints such as mandatory monitoring before inpatient discharge and valid temporal ordering of transitions.  
Conformance checking revealed deviations related to the absence of an ER Visit or where temporal inconsistencies persisted despite correction; no cases of incomplete triage emerged, as triage is represented by atomic events.  
This hybrid procedural–declarative analysis supports interpretability while identifying operational exceptions that warrant review.

\textbf{Step 4. Outcome and Performance Analysis.}
The enriched event log enabled computation of performance indicators and outcome stratifications.  
Metrics include ER and hospital length of stay, daily monitoring intensity (vitals, labs, medications), and discharge disposition by age or ICU exposure.  
These analyses link process variability with clinical results, highlighting points where delays or workload accumulation affect outcomes.

\textbf{Step 5. Managerial Interpretation.}
The combined outputs were synthesized into actionable indicators supporting hospital governance: throughput times, compliance ratios, monitoring intensity, and ICU transition patterns.  
Such metrics can guide triage reliability checks, capacity planning, staffing allocation, and ICU step-down coordination, illustrating the practical relevance of process mining for operational decision making.

\begin{figure}[t]
  \centering
  \includegraphics[width=.70\linewidth]{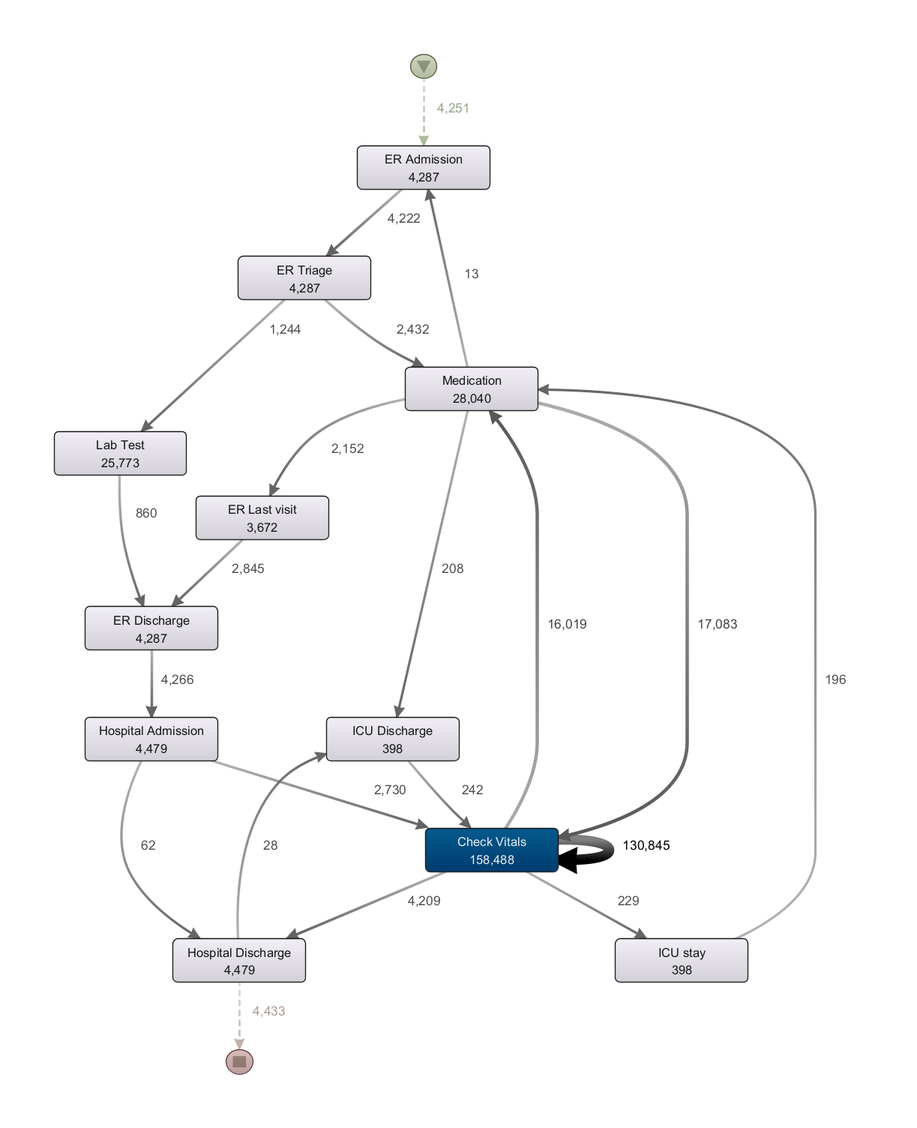}
  \caption{Discovered directly-follows graph (DFG) on CDSL pathways.}
  \label{fig:dfg}
\end{figure}
\begin{figure}[t]
  \centering
  \includegraphics[width=.84\linewidth]{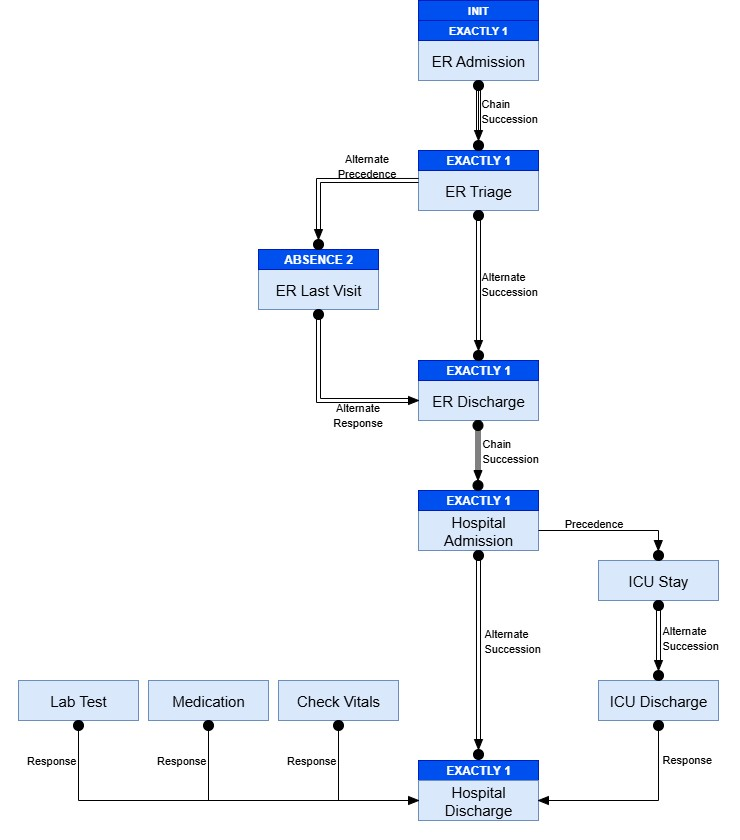}
  \caption{Declarative reference model used for conformance checking.}
  \label{fig:model}
\end{figure}
\begin{figure}[t]
  \centering
  \includegraphics[width=.82\linewidth]{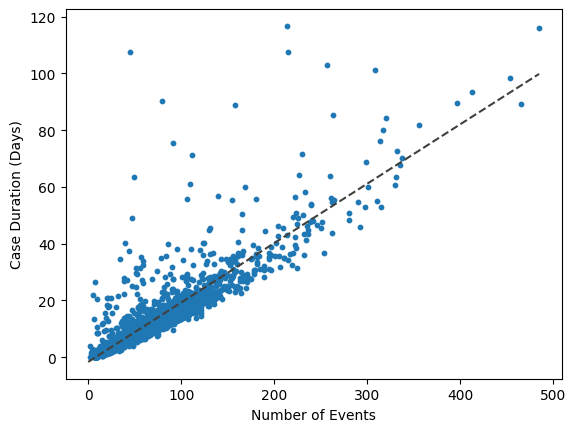}
  \caption{Relation between number of events and case duration (days).}
  \label{fig:evdur}
\end{figure}

\begin{figure}[t]
  \centering
  \includegraphics[width=.80\linewidth]{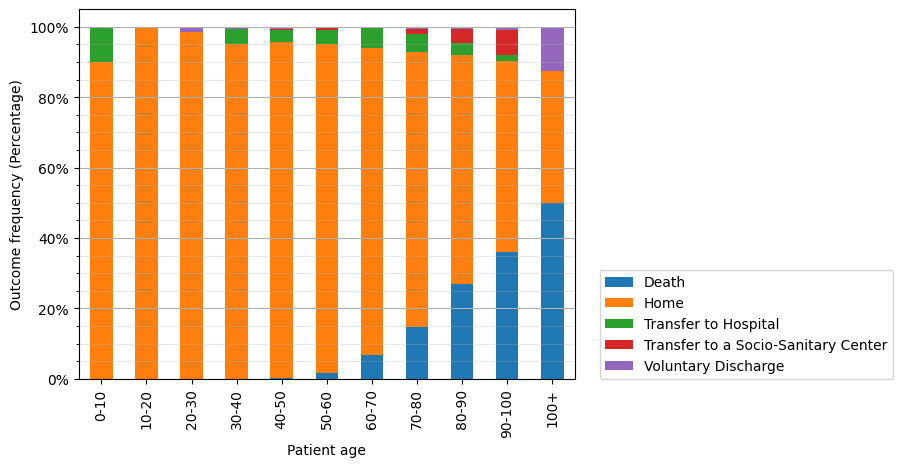}
  \caption{Outcome distribution by patient age. Mortality increases with age, while home discharges decline; transfers concentrate among older cohorts.}
  \label{fig:age}
\end{figure}

\begin{figure}[t]
  \centering
  \includegraphics[width=.80\linewidth]{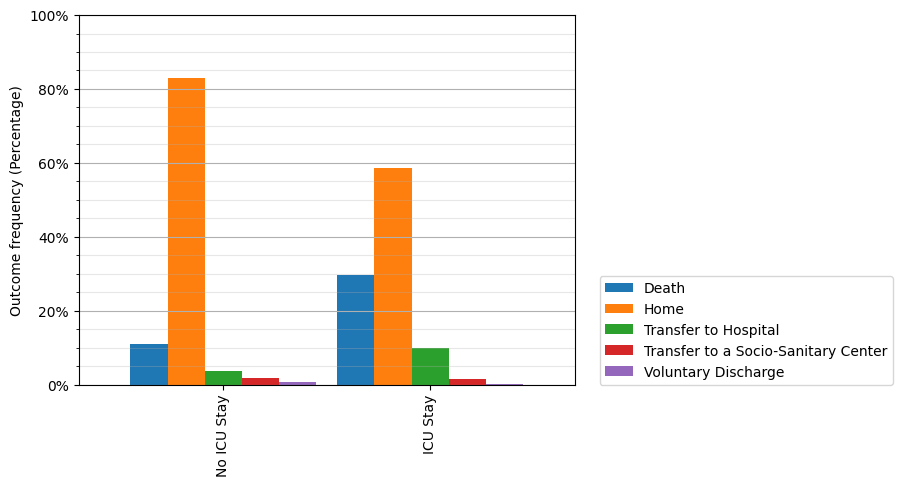}
  \caption{Outcome distribution by ICU stay. ICU episodes show higher mortality and transfers, consistent with longer and more service-intensive trajectories.}
  \label{fig:icu}
\end{figure}

\section{Case Study and Results}
\label{sec:results}
The proposed methodology was applied to the CDSL dataset to demonstrate how the pipeline supports the discovery, conformance, and outcome analysis of hospital processes.  
This section presents the empirical findings in direct relation to the three research questions defined in Section~\ref{sec:intro}:  
\textbf{(RQ1)} What are the dominant trajectories and bottlenecks?  
\textbf{(RQ2)} To what extent do real traces conform to a reference pathway?  
\textbf{(RQ3)} How do process and patient factors relate to outcomes?

\subsection{RQ1. Structural and Temporal Insights}

Figure~\ref{fig:dfg} shows the directly-follows graph (DFG) discovered from the validated event log, with edge labels reporting transition frequencies between activities.  
The model highlights a consistent backbone from \emph{ER Admission} to \emph{Hospital Stay} and \emph{Discharge}, with optional transitions through \emph{ICU Stay}. Monitoring activities, primarily \emph{Check Vitals}, appear as recurrent events throughout the inpatient trajectory, reflecting their role as the routine and repeated component of patient monitoring. Temporal indicators show a median ER length of stay of 6.2 hours (IQR 3.8--10.4) and a median hospital stay of 11.7 days (IQR 7--18). Approximately 8\% of patients experienced an ICU transfer, typically after prolonged ER observation, confirming both the heterogeneity of clinical pathways and the operational pressure on critical care resources during pandemic peaks. As shown later in Figure~\ref{fig:evdur}, the number of recorded events increases approximately linearly with case duration, indicating that monitoring activities tend to be distributed uniformly over time rather than intensifying disproportionately during longer episodes.

\subsection{RQ2. Conformance and Deviations}

Declarative conformance analysis, based on the \emph{Declare} reference model shown in Figure~\ref{fig:model}, revealed high overall adherence to the expected clinical logic encoded in the constraints.  
Rules such as mandatory triage before ER discharge and the requirement that ward discharge follows at least one monitoring event were consistently satisfied across all traces. 
The deviations observed during conformance checking did not concern violations of these core clinical rules. They were mainly related to a complete \emph{ER visit absence}, potentially attributable to either an actual absence or a missed registration, and \emph{timestamp-related inconsistencies}, such as out-of-order event sequences (e.g., monitoring events recorded minutes before an imputed ER arrival time) or duplicate entries with identical timestamps.
These anomalies reflect documentation variability rather than clinical shortcuts, and were addressed during preprocessing through deterministic timestamp correction rather than event removal.
Deviations were more frequent in short ER stays (less than 4 hours), which tend to concentrate borderline or low-information sequences, where minor timestamp shifts produce apparent rule misalignments.  
Although these deviations do not reflect violations of clinical rules, they highlight areas where documentation tends to be less reliable, particularly in short ER stays.

\subsection{RQ3. Outcome Stratifications}
To evaluate how process variability relates to clinical outcomes, we analyzed mortality and discharge patterns by age, ICU exposure, and monitoring intensity (Figures~\ref{fig:age} and~\ref{fig:icu}). Mortality increases steadily with age, especially beyond 70 years, and is markedly higher among ICU-exposed episodes, reflecting greater clinical severity. The relationship between monitoring intensity and outcomes is more nuanced.  
Episodes with moderate monitoring levels (4–5 vital checks per day) show lower mortality and shorter hospital stays, suggesting timely and consistent observation. However, when monitoring intensity exceeds 6 vital checks per day, mortality rises again. This pattern likely reflects a shift in case-mix rather than a causal effect: patients monitored more frequently tend to be clinically unstable and require intensive observation. These stratifications illustrate how process-derived indicators, such as age, ICU exposure, and monitoring frequency, complement traditional risk metrics by adding an operational and temporal perspective to outcome assessment.

\subsection{Operational Implications}
The combined discovery, conformance, and outcome analyses highlight several operational levers for hospital governance:

\begin{itemize}[leftmargin=*,nosep]
  \item \textbf{Triage completeness.} Although not violated in the conformance analysis, periods of timestamp inconsistency and documentation gaps cluster around short ER stays. These patterns can inform targeted improvements in ER documentation practices.
  \item \textbf{Monitoring intensity.} Vital-sign and laboratory frequency serve as proxies for workload and clinical acuity. The non-linear relationship between monitoring intensity and mortality suggests that workload indicators should be interpreted jointly with patient severity markers rather than in isolation.
  \item \textbf{ICU transitions.} The structural variability observed at the ER--ICU interface highlights capacity-sensitive points where delays or prolonged observation are more likely. These insights can guide policies for early escalation, bed allocation, and ICU step-down coordination.
  \item \textbf{Outcome gradients.} Age and severity-related patterns support risk-based staffing strategies and clearer criteria for escalation and discharge planning.
\end{itemize}

Together, these findings demonstrate how a process-mining perspective can integrate structural, temporal, and outcome-related evidence to support continuous improvement in hospital operations.

\section{Discussion}
\label{sec:discussion}
The analyses provide a coherent view of COVID-19 care processes and jointly address the research questions. The pipeline reveals four operational priorities:

\begin{enumerate}[leftmargin=*]

\item \textbf{ER–Admission Handoff}
\emph{Finding:} Timing variability and documentation gaps cluster at ER$\rightarrow$Ward transitions.
\emph{Action:} Standardize the digital handoff; monitor \emph{time from ER arrival to admission}.
\emph{KPIs:} ER dwell time; completeness of ER assessment; ER–ward transfer delay.

\item \textbf{Workload via Event Density}
\emph{Finding:} Event density (events/day) reflects workload and patient acuity.
\emph{Action:} Use density trends to anticipate peaks and adjust staffing or lab throughput.
\emph{KPIs:} vitals-per-day; nurse workload index; lab turnaround time.

\item \textbf{Risk-Stratified Escalation/Discharge}
\emph{Finding:} Outcomes vary by age and ICU exposure; monitoring intensity is non-linear due to case-mix differences.
\emph{Action:} Integrate age, ICU flags, and monitoring profiles into escalation and discharge decisions.
\emph{KPIs:} ICU transfer delay; 72h readmission; discharge-before-noon rate.

\item \textbf{Timestamp Reliability and Data Governance}
\emph{Finding:} Deviations largely result from incomplete or inconsistent timestamps.
\emph{Action:} Enforce timestamp completeness; maintain a stable label set; run routine conformance audits.
\emph{KPIs:} completeness of key timestamps.

\end{enumerate}

To better judge the quality of our work, the following validity threats have to be considered. \textbf{Internal validity.} Analyses rely on routine clinical documentation. Inaccurate or missing timestamps may affect throughput estimates or alter the sequence of temporally close events. \textbf{Construct validity.} Results depend on the chosen activity labels and integration rules. These are explicitly defined, and temporal consistency is verified through systematic quality checks. \textbf{External validity.} The analysis is based on data from a single hospital group, HM Hospitales, during the COVID-19 pandemic. This limits generalizability to other settings or periods. However, the pipeline’s modular design supports adaptation to different institutions. Broader validation across diverse contexts remains a goal for future work. \textbf{Reliability.} Event construction and timestamp alignment follow deterministic, reproducible rules. Corrections are applied only to restore clinical plausibility. \textbf{Information bias.} Some records lack documentation of the ER visit. These omissions likely reflect documentation overload rather than clinical absence and may themselves indicate operational stress.

\section{Conclusions}
\label{sec:conclusion}
This study shows that a reproducible data-to-log pipeline combined with process mining can transform routine hospital data into \emph{management-ready} evidence. Applied to the CDSL dataset, the approach reconstructed ER, inpatient, and ICU pathways, verified adherence to a declarative model, and linked process behavior with outcome patterns. Key findings include a monitoring backbone dominated by \emph{Check Vitals}, documentation inconsistencies at the ER–admission transition, and outcome gradients shaped by age and ICU exposure. 
Future work will extend the analysis integrating predictive monitoring for early warnings and coupling models with what-if simulations to support capacity planning.

\paragraph{Declarations.} We acknowledge financial support under the National Recovery and Resilience Plan (NRRP), M4C2I1.1, funded by the European Union – NextGenerationEU– Project Title aRtificial intElligence for Process Analytics (REPA) - Grant Assignment Decree No. 2022CJWPNA by the Italian Ministry of Ministry of University and Research (MUR).
\bibliographystyle{splncs04}
\bibliography{worldcist_refs.bib}

\end{document}